\documentclass[aps,superscriptaddress,twocolumn,prb]{revtex4-1}

\usepackage{amsmath}
\usepackage{amsfonts}
\usepackage[ansinew]{inputenc}
\usepackage{epsfig}
\usepackage[colorlinks,hyperindex]{hyperref}
\usepackage{graphicx}

\usepackage[separate-uncertainty=true]{siunitx}
\DeclareSIUnit\monlayer{ML}

\hypersetup{
	colorlinks,%
	citecolor=black,%
	linkcolor=black,%
	urlcolor=black,%
}

\newcommand{\mos}{MoS$_2$}
\newcommand{\iv}{d$I$/d$V$}

\makeindex

\begin{document}

\title{Mapping the perturbation potential of metallic and dipolar tips in tunneling spectroscopy on \mos}
\author{Nils Krane}
\author{Christian Lotze}
\email{c.lotze@fu-berlin.de}
\affiliation{Fachbereich Physik, Freie Universit\"at Berlin, Arnimallee 14, 14195 Berlin, Germany.}

\author{Nils Bogdanoff}
\author{Ga\"el Reecht}
\affiliation{Fachbereich Physik, Freie Universit\"at Berlin, Arnimallee 14, 14195 Berlin, Germany.}

\author{Lei Zhang}
\author{Alejandro L. Briseno}
\affiliation{University of Massachusetts, Department of Polymer Science and Engineering, Amherst, USA}

\author{Katharina J. Franke}
\affiliation{Fachbereich Physik, Freie Universit\"at Berlin, Arnimallee 14, 14195 Berlin, Germany.}

\date{\today}

%\keywords{Vibronic states, molecular vibrations, rotamer, BTTT, molybdenum disulfide, \mos, scanning tunneling microscopy}

%--------------------------------------------------------

\begin{abstract}

Scanning tunneling spectroscopy requires the application of a potential difference between the sample and a tip.
In metal--vacuum--metal junctions, one can safely assume that the potential is constant along the metallic substrate.
Here, we show that the inhomogeneous shape of the electric potential has to be taken into account when probing spatially extended molecules on a decoupling layer.
To this end, oligothiophene-based molecules were deposited on a monolayer of molybdenum disulfide (\mos) on a Au(111) surface.
By probing the delocalized molecular orbital along the thiophene-backbone, we found an apparent intramolecular shift of the positive ion resonance, which can be ascribed to a perturbation potential caused by the tip.
Using a simple model for the electrostatic landscape, we show that such a perturbation is caused by the inhomogeneity of the applied bias potential in the junction and may be further modified by an electric dipole of a functionalized tip.
The two effects can be disentangled in tunneling spectra by probing the apparent energy shift of vibronic resonances along the molecular backbone.
We suggest that  extended molecules on \mos\ can be used as a sensor for the shape of the electrostatic potential of arbitrary tips.
\end{abstract}

\maketitle

% ------------ Introduction ----------------

\section{Introduction}
Scanning tunneling microscopy (STM) is frequently employed for the determination of energy levels of individual molecules on surfaces.
Because STM relies on conducting surfaces, metals are the most frequent choice as a substrate for the molecules.
However, the molecular states are strongly modified by hybridization and screening from the metal substrate.
To preserve the character of isolated molecules, thin decoupling layers are inserted between the molecule and the metal.
Prominent examples are thin layers of ionic materials, such as NaCl \cite{ReppPRL2005, GrossPRL2011, VillagomezSurfSci2009, GrillACIE2009} or Al$_2$O$_3$ \cite{QiuPRL2004, LiPRL2015}.
Alternatively, monolayers of molecules  \cite{FrankePRL2008,MatinoPNAS2011}, h-BN \cite{JarvinenNanoLett2013, JoshiACSnano2013, SchulzACSnano2013} or graphene \cite{ChoNanoLett2012, RissACSnano2014, CrommieNatCom2016} have been employed.

The incorporation of thin insulating layers implies a second tunneling barrier between an adsorbate and the metal substrate, such that the molecular states are no longer pinned to the Fermi level of the substrate.
Because of the second tunneling barrier, the molecules do not lie on the same potential as the metallic substrate.
The finite voltage drop across the decoupling layer requires a larger absolute bias voltage for probing the molecular energy levels than without decoupling interlayer \cite{QiuPRL2004,NazinPNAS2005}.
As the fraction of the potential drop across the layer is tunable by the tip--sample distance, it has been suggested to utilize this effect as a gating potential, enabling controlled charging of single molecules \cite{NazinPNAS2005,TorrentePRL2012,HauptmannPCCP2013} or buried impurities\cite{TeichmannPRL2008}.
 Such a gating effect of the tip is also important on semiconducting substrates as band bending shifts all states, causing new quantum well states \cite{DombrowskiPRB1999,FeenstraSurfSci2009,JiangNatNano2017}.

Furthermore, it has been shown that the gating potential does not only depend on the vertical tip--sample distance, but also on the lateral distance from the molecule/impurity \cite{NazinPNAS2005, TeichmannPRL2008, TorrentePRL2012,HauptmannPCCP2013}.
The qualitative distance dependence of the charging peaks could be understood with a point-like tip model with the distance scaling according to simple geometrical considerations.
Hence, the spatial distribution of the potential within both tunneling barriers could be neglected.
We note that the peaks signaling the charging event were of several tens of millielectronvolt (\SI{}{\milli\electronvolt}) width.

Recently, a single layer of \mos\ has been suggested as a decoupling layer, where tunneling through electronic states of molecules revealed sharp resonances of only a few \SI{}{\milli\electronvolt} width \cite{KraneACSnano2018}.
In this work, we show that the molecular resonances exhibit an apparent shift along the molecular backbone by some tens of \SI{}{\milli\electronvolt}.  
At first sight, this is in disagreement with the resonances originating from the same molecular orbital.
However, we explain that this apparent shift occurs as a result of the inhomogeneous potential in the molecular plane.
Hence, we suggest that these narrow resonances are an ideal sensor for mapping the shape of the potential at the position of the molecule within the junction.
To benchmark the sensor, we functionalize the STM tip with a Cl atom yielding an additional dipolar potential.

\begin{figure}[ht]	% ----------------- FIGURE -----------------------
  \begin{center}
    \includegraphics[width=0.8\columnwidth]{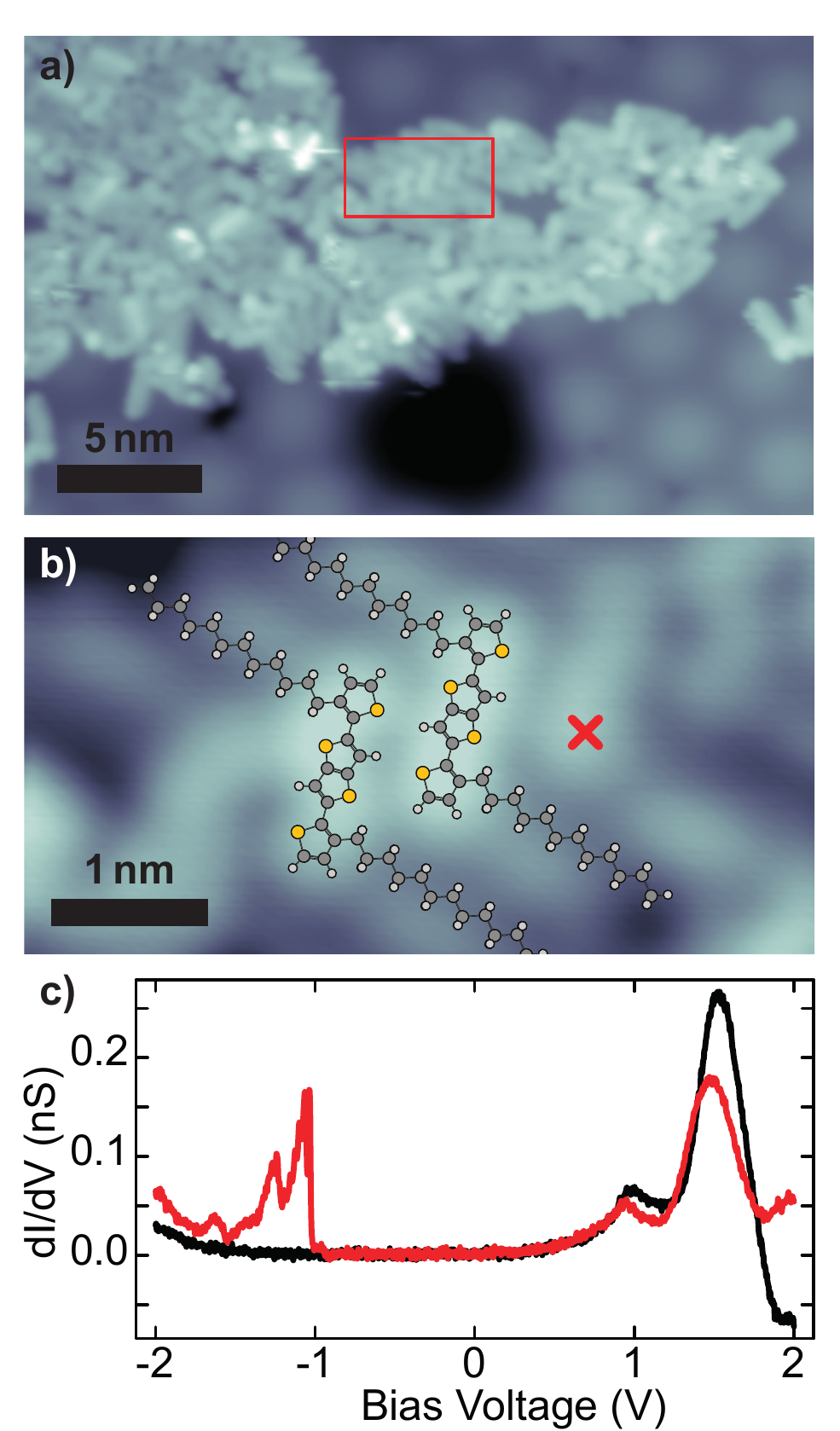}
     \caption{\textbf{a)} Island of BTTT on \mos/Au(111).
The dark area in the bottom is due to a vacancy island in the underlying Au(111) surface \cite{KraneNanoLett2016}(\SI{1}{\volt}/\SI{20}{\pico\ampere}).
\textbf{b)} Zoom into ordered one-dimensional phase with overlayed stick-and-ball models of BTTT molecules (\SI{1}{\volt}/\SI{20}{\pico\ampere}).
The red cross marks the tip position of the \iv\ spectrum taken on BTTT.
\textbf{c)} \iv spectrum of \mos\ (black) and BTTT (red): The PIR and the onset of the NIR of BTTT can be observed at \SI{-1}{\volt} and \SI{2}{\volt}, respectively \cite{KraneACSnano2018}.
(feedback opened at: \SI{2}{\volt}/\SI{100}{\pico\ampere}; $V_\text{rms}=\SI{5}{\milli\volt}$)
}
	\label{fig:01}
	\end{center}
\end{figure}

% ------------ BTTT ----------------
As a model system we use 2,5-bis\-(3-dodecyl\-thiophen-2-yl)thieno[3,2-b]thiophene (BTTT) molecules \cite{ZhangJACS2014} adsorbed on a monolayer of \mos\ on Au(111).
The molecule consists of a thiophene and thienothiophene based backbone with two C$_{12}$H$_{25}$ chains attached to the two terminal thiophene groups (see Fig.~\ref{fig:01}b).

\begin{figure}[ht]	% ----------------- FIGURE -----------------------
	\begin{center}
		\includegraphics[width=0.98\columnwidth]{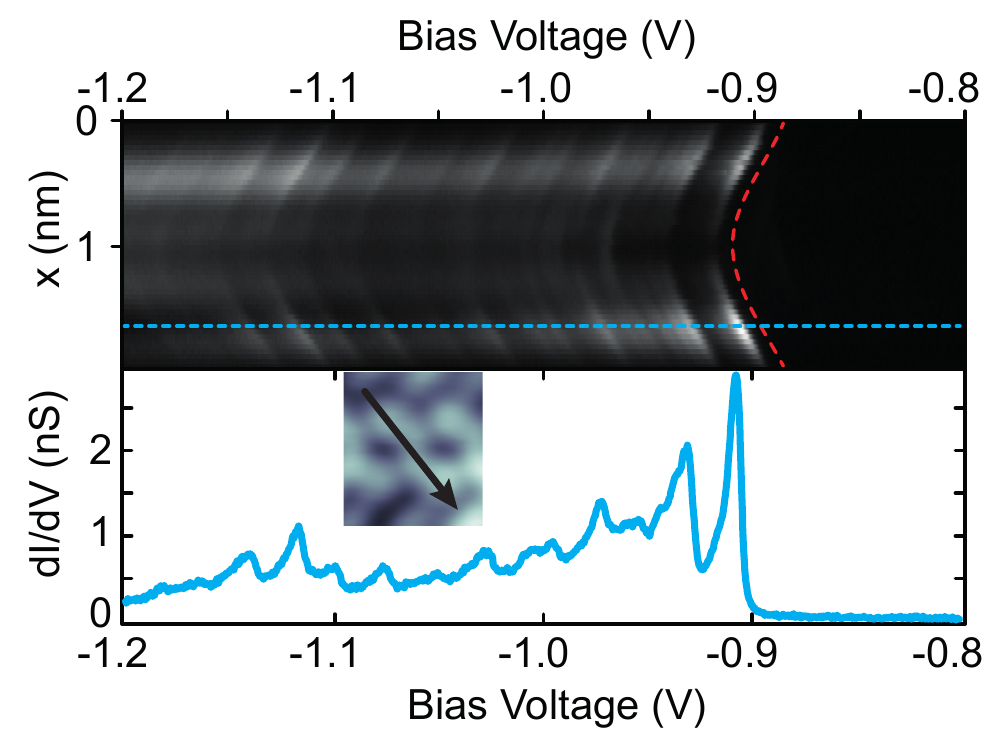}
\caption{Line of 50 \iv\ spectra along the thiophene backbone of BTTT (see inset), displaying the PIR and its vibronic satellites.
The red dashed line highlights the shift of the elastic and inelastic peaks to larger negative bias voltages at the center of the molecule.
A single spectrum taken at the end of the molecule (marked by blue dashed line) is displayed in the bottom panel.
(Feedback opened at the end of the thiophene backbone: \SI{-1.2}{\volt}/\SI{300}{\pico\ampere}; $V_\text{rms}=\SI{0.5}{\milli\volt}$)
}
		\label{fig:02}
	\end{center}
\end{figure}

% ------------ Electronic Structure ----------------
\section{Methods}
A monolayer of \mos\ was grown on a clean Au(111) surface by deposition of Mo and subsequent heating in an H$_2$S atmosphere \cite{SorensenACSnano2014,KraneSurfSci2018}.
BTTT molecules were evaporated from a Knudsen cell evaporator (\SI{365}{\kelvin}) onto the sample held at  \SI{200}{\kelvin}.
All STM measurements were carried out at \SI{4.6}{\kelvin}.
Differential conductance spectra were recorded using a standard lock-in technique (modulation frequency of $f=\SI{921}{\hertz}$).
A metallic gold tip was prepared by controlled indentation of a tungsten tip into the clean Au(111) substrate.
Its quality was regularly assured by reference spectra on Au(111) and \mos.
For a single BTTT molecule in gas phase we performed DFT calculations, using the Gaussian 09 package with the  B3PW91 functional and the 6-31g(d,p) basis set\cite{g09}.

\section{Results and Discussion} 
\subsection{Adsorption structure and electronic resonances of BTTT on \mos}
When evaporated onto a \mos/Au(111) sample held at \SI{200}{\kelvin} the majority of the BTTT molecules form large islands on the Au(111) surface.
A smaller fraction of the molecules forms partially ordered islands on the \mos\ as shown in Fig.~\ref{fig:01}a.
The most common structure in the short-ranged order is a quasi one-dimensional arrangement, where the BTTT molecules are lying with their thiophene backbone and alkyl chains parallel next to each other (Fig.~\ref{fig:01}b).

To investigate the electronic properties of BTTT on \mos\ we performed constant-height tunneling spectroscopy (see Fig.~\ref{fig:01}c).
The reference spectrum on \mos\ shows the typical conduction band states at \SI{0.9}{\volt} and \SI{1.4}{\volt}, as well as the onset of the valence band at \SI{-2}{\volt} \cite{KraneSurfSci2018}.
The spectrum taken on a BTTT molecule shows a narrow positive ion resonance (PIR) at \SI{-1}{\volt}, which can be ascribed to tunneling through the highest occupied molecular orbital (HOMO).
The negative ion resonance (NIR) reflecting tunneling through the lowest unoccupied molecular orbital (LUMO) lies outside of the \mos\ gap ($>$ \SI{2}{\volt}) \cite{KraneACSnano2018}.
Because the PIR is located inside the electronic gap of \mos, it is electronically decoupled from the environment and exhibits a very narrow line width of \SI{6}{\milli\electronvolt}.
The narrow line width allows for the observation of vibronic satellite peaks at larger negative bias voltages.
These were analyzed in more detail in a previous work \cite{KraneACSnano2018}.

To probe the effect of the tip's location on the measured electronic resonances, we plot a set of 50 spectra taken along the BTTT backbone in Fig.~\ref{fig:02}.
The onset of tunneling through the HOMO is shifted by $\sim$\SI{25}{\milli\volt} to larger negative bias voltages in the center of the molecule than at its terminations (see red dashed line as guide to the eye).
The set of vibronic peaks follows the same trend.
We note that the strongest \iv\ signal is observed at the terminations of the thiophene unit with less signal in the center, even though the HOMO is delocalized along the thiophene backbone.
This intensity distribution can be simulated employing an s-wave tip and calculating the tunneling matrix element using Bardeen's approach \cite{BardeenPRL1961} at a certain height above the molecule (see Appendix~\ref{App:Bardeen}).

\begin{figure}[th]	% ----------------- FIGURE -----------------------
   \begin{center}
    \includegraphics[width=0.98\columnwidth]{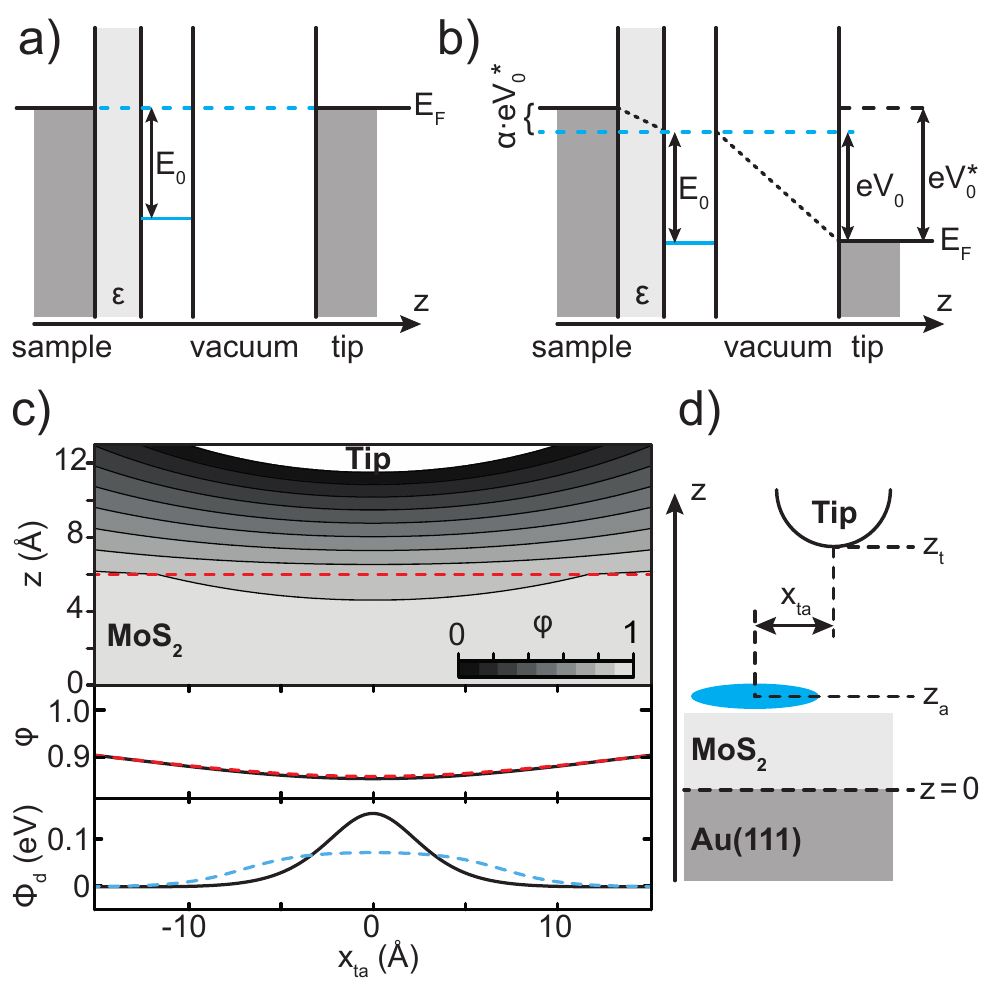}
    \caption{
    \textbf{a,b)} Plate capacitor model of the tunneling junction without (a) and with (b) applied bias voltage.
    The solid blue line represents the electronic state of an adsorbate.
    \textbf{c)} \textit{Top Panel:}
    Simple two charge model of $\varphi(x,z)$ in the tunnel junction (see Appendix), with tip radius $R=\SI{20}{\angstrom}$, and tip--sample distance $z_\text{t}=\SI{6.5}{\angstrom}$.
    The effect of the dielectric constant of \mos\ was approximated by stretching the vertical axis within the \mos\ layer (below the dahed red line) by a factor of 6.
    \textit{Mid Panel:}
    Shape of $\varphi(x_\text{ta},z_\text{a})$ (black solid) above the \mos\ layer (along the dashed line in top panel) as well as effective bias potential $\langle \phi_b(x_\text{ta},z_\text{a}) \rangle$ obtained by perturbation theory (dashed red line).
    \textit{Bottom Panel:}
    Electric potential $\phi_d(x_\text{ta},z_\text{a})$ (black) due to a dipole located at the tip with positive charge pointing towards the surface ($p=-3\,\text{D}$) and distance $d=\SI{3}{\angstrom}$ between the charges.
    The effective dipole potential $\langle \phi_d(x_\text{ta},z_\text{a}) \rangle$ from perturbation theory is plotted by the dashed blue line.
    \textbf{d)} Sketch of the the junction and the used labels.
}
    \label{fig:03}
  \end{center}
\end{figure}

\subsection{Spatial variation of molecular resonances due to an inhomogeneous tip--substrate potential} 
To understand the apparent intramolecular shift of the PIR, we have to take the shape of the electric potential in the STM junction into account.
Because the decoupling layer acts as an insulator, the junction can be described as a double tunneling barrier.
We consider an electronic state of an adsorbate between these barriers with an energy of $E_0$ with respect to the aligned Fermi levels of the two leads (Fig.~\ref{fig:03}a).
When a bias voltage $V_b$ is applied between surface and tip, it drops over both barriers as depicted in Fig.~\ref{fig:03}b.
Hence the effective bias voltage, applied to the adsorbate, is reduced by a factor of $\alpha \leq 1$.
In order to measure the adsorbate`s state, the voltage drop across the decoupling layer needs to be compensated and the bias voltage has to be increased to $V>E_0/e$.
In fact, the adsorbate state appears in a \iv-spectrum at a bias voltage of\cite{QiuPRL2004, NazinPNAS2005}
\begin{equation}
 V^\ast_0 = \frac{E_0}{e \left (1-\alpha \right )} = \frac{V_0}{1-\alpha}.
\end{equation}

In the often used model of a plate capacitor for the tunneling junction, the potential is constant within a plane parallel to the substrate.
The factor $\alpha$ is therefore independent of the lateral distance between tip and adsorbate.
For a more realistic tip model, however, the electrostatic potential at a certain point in the junction depends not only on the vertical but also on the lateral distance to the tip.
A position-independent factor $\alpha$ is no longer sufficient to account for the voltage drop across the decoupling layer.

In first order, the inhomogeneous electric potential in the junction scales with the applied bias voltage and depends on the position.
% It can be described by $\phi_b(\vec r)=V_b \cdot \varphi(\vec r)$, with $\varphi(\vec r)$ being a position-dependent scalar and $V_b$ the applied bias voltage.
It can be described by $\phi_b(\vec r)=V_b \cdot \varphi(\vec r)$, with $\varphi(\vec r)$ being the potential shape of dimension unity and $V_b$ the applied bias voltage.
As the tip is grounded, the potential vanishes there ($\phi_b(\mathrm{tip})=0$), whereas at the metal surface it is equal to the bias voltage ($\phi_b(\mathrm{metal\, substrate})=V_b$). %\cite{KocicJCP2017}
The voltage between an adsorbate at position $\vec r_\text{a}$ and the tip is then $V_b \cdot \varphi(\vec r_\text{a})$.
Accordingly, the voltage drop across the decoupling layer is $V_b \cdot (1-\varphi(\vec r_\text{a}) )=V_b \cdot \alpha(\vec r_\text{a})$.
Here we approximate a vanishing contact potential difference (CPD), which would introduce a bias offset \cite{KocicJCP2017} (see Appendix~\ref{App:CPD}).

Due to the position-dependent voltage drop, the energy level of the adsorbate can be tuned with respect to the sample, which has been employed for an effective gating of the adsorbate state \cite{NazinPNAS2005,TeichmannPRL2008,TorrentePRL2012}.
It depends on the distance of the tip from the molecule, both in the lateral as well as vertical direction.
As a result charging rings appear when the adsorbate`s state crosses the Fermi level of the substrate.

In the following, we explain that the shape $\varphi(\vec r)$ of the potential in the tunneling barrier is responsible for the aforementioned apparent shift of the molecular resonance across the molecule.
For this we use a simple model to describe $\varphi(\vec{r})$  between tip apex and sample.
The tip is approximated by a point charge and the constant electric potential of the sample surface is ensured by a mirror charge of opposite sign in the metal substrate.
A more detailed description of the model is given in the Appendix.
% The resulting $\varphi(x,z)$ is represented by several equipotential lines in the top panel of Fig.~\ref{fig:03}c.
The resulting $\varphi(x,z)$ is represented by several contour lines in the top panel of Fig.~\ref{fig:03}c.
In accordance with our experimental setup the tip is grounded ($\varphi (\mathrm{tip})=0$) and the bias voltage (here normalized to 1) is applied to the surface ($\varphi(x,0)=1$).

The resulting potential shape at the height $z_\text{a}$ of an adsorbate on the decoupling layer $\varphi(x_\text{ta},z_\text{a})$  is shown in the mid panel of Fig.~\ref{fig:03}c (solid black line) with $x_\text{ta}$ being the lateral distance between tip and adsorbate.
% The resulting potential at the height $z_\text{a}$ of an adsorbate on the decoupling layer $\varphi(x_\text{ta},z_\text{a})$  is shown in the mid panel of Fig.~\ref{fig:03}c (solid black line) with $x_\text{ta}$ being the lateral distance between tip and adsorbate.
In order to account for the dielectric constant $\epsilon_\text{lay}$ of the \mos\ layer, we approximated $z_\text{a} = d_\text{lay}/\epsilon_\text{lay} =\SI{6.1}{\angstrom}/6.4 \sim \SI{1}{\angstrom}$, with $d_\text{lay}$ being the thickness of a \mos\ monolayer \cite{LaturiaNPJ2018}.
Because of the curvature of the tip apex, $\varphi(x_\text{ta},z_\text{a})$ is minimal when the tip is placed right above the adsorbate ($x_\text{ta}=0$).

To analyze the effect of the inhomogeneous potential on molecular states of an adsorbate, we employ a simple perturbation model (for details see Appendix).
We represent the molecular orbital of the BTTT by the seventh mode of a particle in a one-dimensional box (PiB) of length \SI{14}{\angstrom}.
The perturbation potential within the box is given by the inhomogeneous potential along the molecule as determined in Fig.~\ref{fig:03}c (solid black line).
The effective potential shape $\langle \varphi(x_\text{ta}) \rangle$ of the PiB was calculated in first order perturbation theory and displayed in Fig.~\ref{fig:03}c (dashed red line).
In this simple model, the effective potential can be roughly approximated as an average of the bias potential along the molecule.
Since $\langle \varphi(x_\text{ta}) \rangle$ depends on the tip position, the required bias voltage to measure a delocalized state ($V^\ast_0 = E_0 / e\langle \varphi(x_\text{ta}) \rangle$) depends on $x_\text{ta}$.
Hence, the molecular energy levels appear to lie at different energies along the molecule, with the shift being strongest in the order of a few percent when the tip is located at the center of the molecule.
This is in agreement with our experimental observation of a larger negative bias voltage being required for probing the PIR in the molecule's center (Fig~\ref{fig:02}).

\subsection{Inhomogeneous potential due to a tip dipole} 
In addition to the inherent inhomogeneous potential in the tip--substrate junction, the tip may carry a local dipole moment at its apex.
STM tips are known to provide a dipole moment of up to a few Debye \cite{GrossPRB2014,GaoACSnano2014}, depending on their termination.
Hence, the applied potential to an adsorbate on a decoupling layer would be a superposition of the potential set by the applied bias voltage $\phi_b$ and the dipole potential $\phi_d$.
The effect of an additional potential in a tunneling junction has been utilized in a slightly different configuration for scanning quantum dot microscopy \cite{WagnerPRL2015}.
Considering the effect of the dipole in a similar perturbation model as above leads to a shift of the electronic energy levels in the \iv\ spectrum to $V^\ast_0 = E_0/e - \phi_d$ (now neglecting the effect of the inhomogeneity of the applied bias voltage).
For a metal tip, the dipole usually points with the positive charge towards the sample, i.e. $\phi_d > 0$ within the junction, thus requiring a larger negative bias voltage for measuring an electronic state in the \iv\ spectrum.
The effective potential of a tip dipole also depends strongly on the lateral distance between tip and adsorbate, as shown in the bottom panel of Fig.~\ref{fig:03}c (solid black).
Employing an equivalent perturbation model as above for the effect of a dipole potential (see Appendix) on a PiB state we obtain the effective potential $\langle \phi_d(x_\text{ta})\rangle$, shown as the dashed blue line across the molecule.
For the measurement of the PIR of BTTT with a metal tip, this would also cause an apparent shift of the state to more negative bias voltages in the center of the molecule.
Hence, a dipole of a metal tip qualitatively shifts the states in the same direction as the inhomogeneity of the potential in the tip--sample junction itself.

\subsection{Comparison of Metal and Cl Tip}
\begin{figure}[th]	% ----------------- FIGURE -----------------------
  \begin{center}
    \includegraphics[width=0.98\columnwidth]{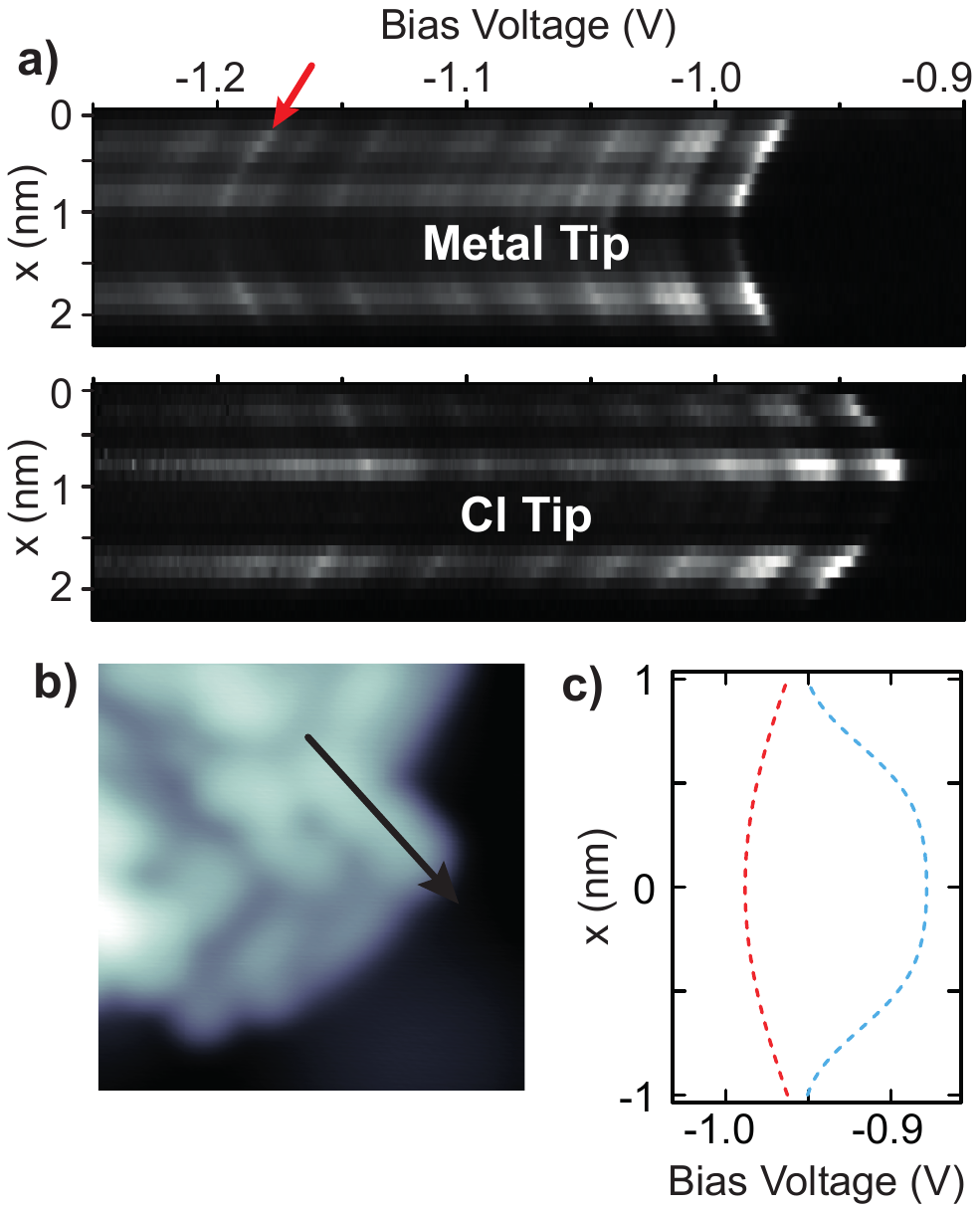}
    \caption{
    \textbf{a)} Line of 22 \iv\ spectra along a BTTT molecule (marked in b) measured with a metal (top) and a Cl functionalized (bottom) tip.
    While the peaks for the metal tip shift to more negative bias voltages, the Cl tip causes a shift to more positive bias voltages in the center of the molecule.
    The red arrow marks the vibronic state of the C--C stretching mode.
    (Feedback opened at the end of the thiophene backbone: \SI{-1.3}{\volt}/\SI{150}{\pico\ampere}; $V_\text{rms}=\SI{1}{\milli\volt}$)
    \textbf{b)} Topography of BTTT molecule, with line of spectra marked by black arrow (\SI{1}{\volt}/\SI{20}{\pico\ampere})
    \textbf{c)} Simulated shift of an electronic state with energy \SI{-0.85}{\electronvolt} in an \iv\ spectrum due to a decoupling layer without (red) and with tip dipole of $p=3\,\text{D}$ (blue).
}
    \label{fig:04}
  \end{center}
\end{figure}

To illustrate the shift originating from the inhomogeneity of the junction and a possible tip dipole, we impose a tip dipole in the opposite direction on the tip.
To this end, we pick up a Cl atom from co-deposited Fe-octaethyl-chloride molecules.
The Cl is expected to be negatively charged at the tip apex, and, thus, exhibit a dipole oriented away from the substrate \cite{GrossPRB2014}.
To exclude any effects of the molecular environment, we first probe the shift of the PIR along a BTTT molecule as shown in Fig.~\ref{fig:04}a.
Afterward, we measure on the very same molecule again with a Cl functionalized tip.
Now, we observe an inverted trend of the shift of all resonances along the molecule.
A smaller negative bias voltage is required for PIR excitation in the center of the molecule than at its terminations.
(The asymmetric intensity in the map is probably due to interaction with the neighboring molecule.)
The inversion of the trend directly reflects that the dipole plays a significant role.
At first sight, it may be surprising that the shift with both tips - though of opposite direction - is of approximately the same size.
The dipole of the metal tip is expected to be much smaller than of the Cl tip \cite{GrossPRB2014}.
However, the effect of the inhomogeneous junction geometry and a metal dipole add up, whereas the dipole of the Cl tip overcompensates the potential shift due to the junction geometry.

Adding up the effect of an inhomogeneous junction and of a tip dipole, the effective electric potential applied to the adsorbate is
\begin{equation}
\langle \phi_\text{a} (x_\text{ta}) \rangle= V_b \cdot \langle \varphi(x_\text{ta})\rangle + \langle\phi_d(x_\text{ta})\rangle.
\end{equation}
It consist of the bias-voltage-dependent term due to the position-dependent voltage drop, as well as a bias-voltage-independent term from the dipole potential.
In accordance, an electronic state with energy $E_0$ is observed in the \iv\ spectrum at a bias voltage of
\begin{equation}
 V^\ast_0(x_\text{ta}) = \frac{E_0/e-\langle\phi_d(x_\text{ta})\rangle}{\langle \varphi(x_\text{ta})\rangle}.
 \label{eq:01}
\end{equation}
Fig.~\ref{fig:04}c displays the calculated $V^\ast_0(x_\text{ta})$ of a decoupled state at $E_0=\SI{-0.85}{\electronvolt}$ for a tip without dipole moment (red) and with dipole moment of $p=3\,\text{D}$ (blue).
With these parameters we can reproduce the observed behavior of the measured data displayed in Fig.~\ref{fig:04}a with our simple model.

In order to disentangle the two origins for both metal and Cl functionalized tip, we utilize the shift of the vibronic resonances as compared to the elastic resonance.
As example, we choose the vibronic resonance associated to the C--C stretching mode at about $\Delta V^\ast = \SI{210}{\milli\volt}$ (marked by red arrow in Fig~\ref{fig:04}a).

\begin{figure}[ht]	% ----------------- FIGURE -----------------------
  \begin{center}
    \includegraphics[width=0.85\columnwidth]{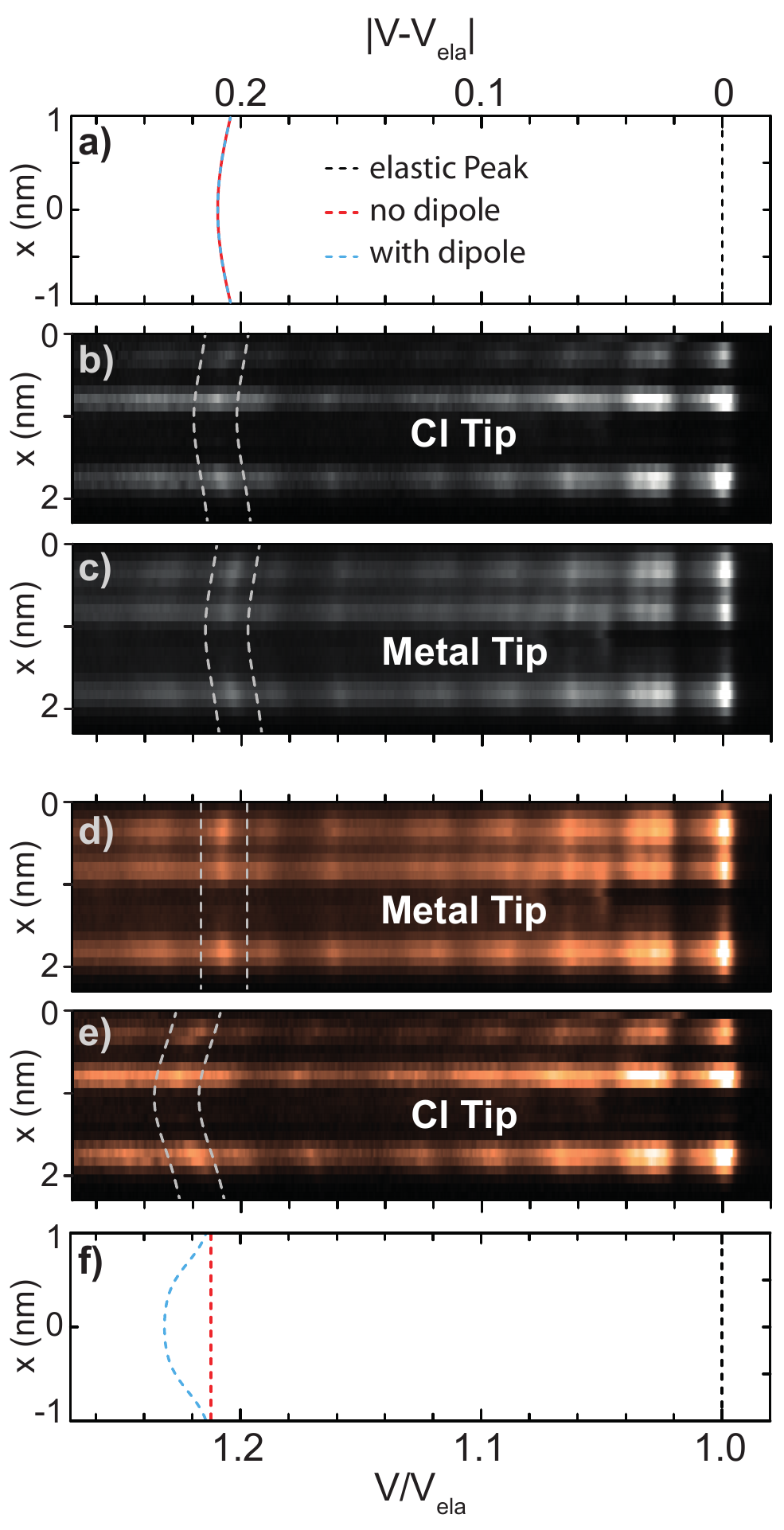}
    \caption{
    \textbf{a)} Calculated shifted bias voltage of elastic (black) and vibronic peak along a molecule for a tip with (blue) and without (red) tip dipole.
    The effect of the constant dipole potential cancels out, yielding an identical shift for both tips.
    \textbf{b,c)} Measured \iv spectra with shifted bias voltage axis for Cl and metal tip, with white dotted lines as a guide for the eye.
    Both tips show a similar shift of the inelastic peak, indicating a similar effect due to the inhomogeneous bias potential.
    \textbf{d,e)} \iv spectra with normalized bias voltage axis for metal and Cl tip.
    For the metal tip, the inelastic peak is nearly constant, pointing towards a negligible tip dipole moment.
    \textbf{f)} Calculated normalized bias voltage of elastic (black) and vibronic peak for a tip with (blue) and without (red) tip dipole.
}
    \label{fig:05}
  \end{center}
\end{figure}

The voltage difference $\Delta V^\ast(x_\text{ta})$ between the elastic and the inelastic peak is accordingly
\begin{equation}
 \Delta V^\ast (x_\text{ta}) =\left | V^\ast_\text{inel.} - V^\ast_\text{ela.} \right |
 = \frac{\hbar\omega}{e\langle \varphi(x_\text{ta})\rangle},
\label{eq:02}
\end{equation}
with $\hbar\omega$ being the energy of the vibrational mode.
Please note, that $\Delta V^\ast$ depends solely on the spatially varying bias potential shape $\langle \varphi(x_\text{ta})\rangle$ and not on the potential of the tip dipole.
In Fig.~\ref{fig:05}a the calculated elastic (black) and the inelastic peaks are plotted along the molecule for a tip without (red) and with dipole (blue).
The energy scale is shifted by $V_\text{ela.}(x_\text{ta})$, thus the elastic peak is always at \SI{0}{\volt}.
As a vibrational energy we use $\hbar\omega=\SI{180}{\milli\electronvolt}$ to account for the 10-\SI{15}{\percent} voltage drop in our model and to reproduce the experimental values.
The voltage difference $\Delta V^\ast(x_\text{ta})$ is largest in the center of the molecule, where $\langle \varphi(x_\text{ta})\rangle$ has a minimum.
Since the dipole potential is canceled out by the subtraction, the inelastic peaks shift independently of the tip dipole.

In contrast, by normalizing the bias voltages with $V^\ast_\text{ela.}(x_\text{ta})$, the term $\langle \varphi(x_\text{ta}\rangle)$ cancels out and we obtain the normalized voltage
\begin{align}
 v(x_\text{ta}) = \frac{V^\ast_\text{inel.}}{V^\ast_\text{ela.}} =  1+\left |\frac{\hbar\omega}{E_\text{ela.}-\langle\phi_d(x_\text{ta})\rangle}\right |
\label{eq:03}
\end{align}
In the case of no tip dipole, $\langle\phi_d(x_\text{ta})\rangle$ vanishes and $v$ is independent of the position of the tip.
Hence, a vibronic state is expected to appear at a constant normalized voltage as displayed by the red dashed line in Fig.~\ref{fig:05}f.
For a tip including a dipole moment, the normalized energy of the vibronic state changes in dependence of the tip position.
When measuring a PIR ($E_\text{ela.} < 0$) with a Cl functionalized tip ($\phi_d(x_\text{ta}) < 0$) the normalized voltage of the vibronic state should shift to a higher normalized voltage at the center of the molecule (blue dashed line in Fig.~\ref{fig:05}f).

The \iv\ spectra measured with both tips along the BTTT are displayed in Fig.~\ref{fig:05} with shifted (b,c) and normalized (d,e) voltage axis.
The vibronic states of the C--C stretch mode are marked by the dashed white lines as a guide for the eye.
In the case of a shifted voltage axis, both tips feature a similar shift of the vibronic state to more negative energies by a few percent, i.e., larger voltage difference between elastic and inelastic peak.
In accordance with the calculation, this confirms the effect of the geometry-imposed inhomogeneous potential for both tips.
Please note that the exact height for both tips are unknown, thus the experimental data can only be compared qualitatively between the two tips and the simulation.

\label{par:results}
For the normalized voltage axis the two tips behave differently.
For the metal tip, the vibronic state is at nearly constant normalized voltage along the molecule, independent of the tip position.
This implies a negligible dipole moment of the metal tip, when a vanishing CPD is assumed (see Appendix~\ref{App:CPD}).
Accordingly the energy shift along the molecule measured by a metal tip is almost exclusively due to the inhomogeneous potential within the junction.
For the Cl tip, the vibronic state shifts strongly to higher normalized voltages as predicted and plotted in Fig.~\ref{fig:05}f for a non-zero dipole moment.

\section{Conclusion}
In conclusion, we presented the perturbation effect of an inhomogeneous electric potential in the STM junction on a delocalized molecular orbital.
Utilizing the excellent decoupling properties of single-layer \mos, we demonstrated that this effect causes an apparent intramolecular shift of the PIR of the thiophene based BTTT molecule.
The inhomogeneous potential in the junction does not only originate from the applied bias voltage, but also from a tip dipole of a functionalized tip.
Using the vibronic resonances in the tunneling spectra we were able to distinguish between the two origins.
This observation is important to take into account, because it is relevant for all adsorbates with delocalized electronic states on a decoupling layer.
In turn, we suggest to use the resonances' shift as a sensor for inhomogeneous potentials and dipole components in the tip.

\appendix

\section{\label{App:Bardeen}Tunneling Probability into PIR}
\begin{figure}
	\includegraphics[width=0.95\columnwidth]{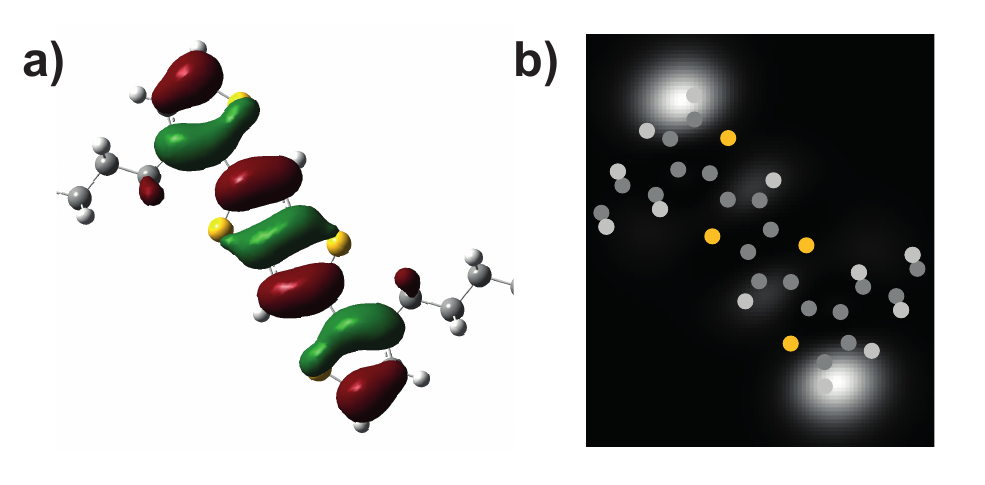}
	\caption{\textbf{a)} Highest occupied molecular orbital of the BTTT molecule.
	\textbf{b)} Calculated constant-height image of the HOMO, measured with an s-wave tip.
	}
	\label{SIfig:03}
\end{figure}

The BTTT molecule was calculated in gas phase, using the Gaussian 09 package with the  B3PW91 functional and the 6-31g(d,p) basis set\cite{g09}.
The constant-height \iv\ image (Fig.~\ref{SIfig:03}b) was then simulated by calculating the tunneling matrix element at each point of the flat-lying molecule \cite{BardeenPRL1961}
\begin{equation}
 M_\text{ta} \propto \int  \text{d}\textbf{S} \left ( \Psi^\ast_\text{t} \nabla \Psi_\text{a} - \Psi_\text{a} \nabla \Psi^\ast_\text{a} \right ).
\end{equation}
Here $\Psi_a$ is the wave function of the molecule obtained by DFT calculations (Fig.~\ref{SIfig:03}a) and $\Psi_t$ the spherical wave function of an s-wave tip.
We choose our integration plane at \SI{1.5}{\angstrom} above the center of the molecule. The calculations reveal the largest tunneling probability at the terminations of the thiophene backbone. The images resemble the experimental data at negative bias voltage, thus evidencing tunneling via the HOMO. 

\section{\label{App:Junction}Junction Model}

\begin{figure*}
	\includegraphics[width=0.7\textwidth]{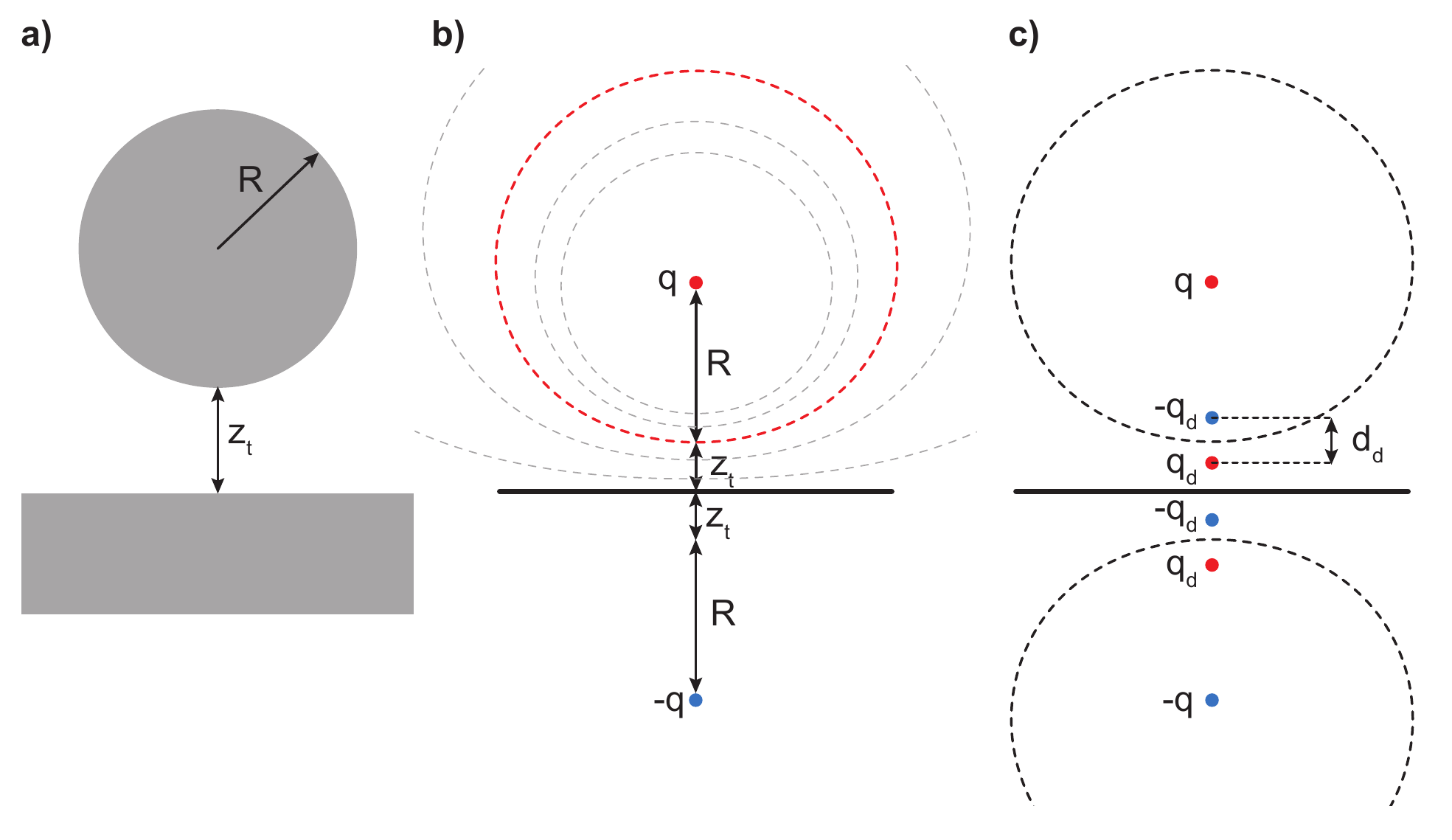}
	\caption{Tip Models:
	\textbf{a)} Spherical tip at distance $z_t$ to the substrate.
	\textbf{b)} Approximating the electric potential by two charges. The dashed lines represent isopotential surfaces of the dipole.
	The isopotential surface at distance $R$ between charge and surface (red) is chosen as a model for the metal tip.
	\textbf{c)} Approximating the tip dipole with two charges $q_d$ and their mirror charges.
	}
	\label{SIfig:01}
\end{figure*}

We model the STM junction as an almost spherical, grounded tip with radius $R$ and distance $z_t$ to a conducting surface, held at bias voltage $V_b$, as shown in Fig.~\ref{SIfig:01}a.
The potential within this junction can be approximated by a point charge $q$ and its mirror charge $-q$ separated by $2(R+z_t)$ at $x=0$ (Fig.~\ref{SIfig:01}b).
The resulting potential reads:
\begin{equation}
\begin{split}
 \phi_b(x,z) = \frac{1}{4\pi\epsilon_0} \left [ \frac{q}{\sqrt{x^2 + (z-R-z_t)^2}} \right . \\
 \left . - \frac{q}{\sqrt{x^2 + (z+R+z_t)^2}}\right ] + C.
\end{split}
  \label{eq:bias}
 \end{equation}
 
We choose the constant potential $C$ in a way, that $\phi_b(0,z_t) = 0$, yielding:
\begin{equation}
\begin{split}
 \phi_b(x,z) = \frac{q}{4\pi\epsilon_0} \left [ \frac{1}{\sqrt{x^2 + (z-R-z_t)^2}} \right . \\
 \left . - \frac{1}{\sqrt{x^2 + (z+R+z_t)^2}} - \frac{1}{\lvert R \rvert} + \frac{1}{\lvert R + 2z_t \rvert} \right ].
 \end{split}
 \label{eq:offset}
\end{equation}
Thus the isopotential surface around the charge $q$ approximates the grounded tip with radius $R$ (dashed red line in Fig.\ref{SIfig:01}b).

The charge $q$ is chosen to cause a certain bias voltage $V_b$ applied to the sample ($\phi_b(x,0) = V_b$):
\begin{equation}
 q = - 4 \pi\epsilon_0 \left ( \frac{R^2}{2z_t} + R \right ) \cdot V_b
 \label{eq:charge}
\end{equation}

Furthermore we can write the potential between surface and tip as
\begin{equation}
 \phi_b(x,z) = V_b \cdot \varphi(x,z).
\end{equation}

Inserting Eq.~\ref{eq:offset} and \ref{eq:charge} yields:
\begin{equation}
\begin{split}
 \varphi(x,z) =  1 - \left ( \frac{R^2}{2z_t} + R \right ) \left [  \frac{1}{\sqrt{x^2 + (z-R-z_t)^2}} \right . \\
 \left . - \frac{1}{\sqrt{x^2 + (z+R+z_t)^2}} \right ].
 \end{split}
 \label{eq:varphi}
\end{equation}
A map of $\varphi(x,z)$ is displayed in Fig.~3 of the main text, for $R=\SI{20}{\angstrom}$ and $z_t=\SI{6.5}{\angstrom}$.

To account for the decoupling layer, we assume the molecule to be at $z_\text{a}=\SI{1}{\angstrom}$ above the substrate.

\section{\label{App:Dipole}Dipole Model}
We approximate the tip dipole with two mirror charges centered to the tip apex as depicted in Fig.~\ref{SIfig:01}c.
For a tip with a dipole $p$ and length $d$, the potential is
\begin{equation}
 \begin{split}
 \phi_d(x,z) = \frac{p/d}{4\pi\epsilon_0} \left [ \frac{1}{\sqrt{x^2 + (z-z_t + d/2)^2}} \right. \\
 \left. - \frac{1}{\sqrt{x^2 + (z-z_t - d/2)^2}}   \right. \\
 \left. - \frac{1}{\sqrt{x^2 + (z+z_t - d/2)^2}} \right. \\
 \left. + \frac{1}{\sqrt{x^2 + (z+z_t + d/2)^2}} \right ]
 \end{split}
 \label{eq:dipole}
 \end{equation}
The potential at $z_\text{a}=\SI{1}{\angstrom}$ is displayed in Fig.~3 of the main text, for $p=-3\,\text{D}$ and $d=\SI{3}{\angstrom}$.

\section{\label{App:pertubation}Perturbation Model}

\begin{figure}
	\includegraphics[width=0.8\columnwidth]{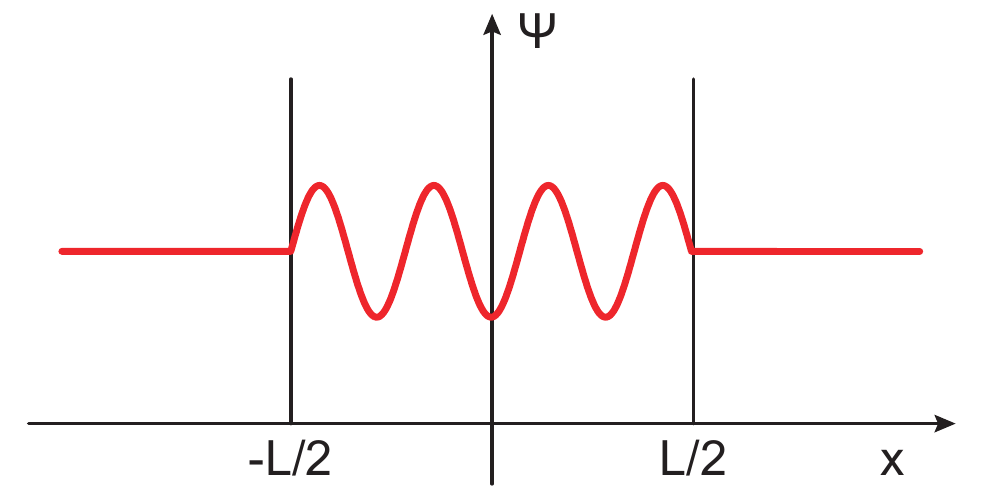}
	\caption{Wave function of the seventh mode of a particle in a one-dimensional potential well with infite walls.
	}
	\label{SIfig:02}
\end{figure}

In order to investigate the influence of an inhomogenous potential to the energy of a molecular state, we use a simple perturbation model.
The molecular orbital is represented by a particle in a one-dimensional box of length $L$.
We choose the seventh mode, to emulate the six nodal planes of the HOMO of the BTTT molecule.
Hence, the wavefunction is described by
$$ \Psi(x)= \begin{cases}
 0 & x < -\frac{L}{2} \\
 \sqrt{\frac{2}{L}} \sin \frac{7\pi}{L}\left( x + \frac{L}{2}\right) \quad &  -\frac{L}{2} < x < \frac{L}{2} \\
 0 &  x > -\frac{L}{2}
 \end{cases}$$
as depicted in Fig.~\ref{SIfig:02}.

The energy of the molecular state within a potential is then shifted by
\begin{equation}
 E (x_\text{ta}) = e \int_{-\infty}^{+\infty} \Psi^*(x) \phi_0(x-x_\text{ta}) \Psi(x)\ \text{d}x,
\end{equation}
whereas $e$ is the elementary charge and $x_\text{ta}$ the lateral position of the tip relative to the molecule.

For the inhomogenous potential in the molecular plane we use:
\begin{align}
 \phi_0(x) &= \phi_b(x) + \phi_d(x) \\
 &= V_b \cdot \varphi(x) + \phi_d(x),
 \label{eq:potential}
\end{align}
with $\phi_b(x)$ and $ \phi_d(x)$ being the bias and the dipole potential, respectively, at height $z_\text{a}$ as described above.

Hence, the shift due to the perturbation is:
\begin{align*}
 E (x_\text{ta}) &= e V_b \int_{-\infty}^{+\infty} \Psi^*(x) \varphi(x-x_\text{ta}) \Psi(x)\ \text{d}x  \\
 &+ e \int_{-\infty}^{+\infty} \Psi^*(x) \phi_d(x-x_\text{ta}) \Psi(x)\ \text{d}x \\
 &= eV_b \langle \varphi(x_\text{ta}) \rangle + e \langle \phi_d(x_\text{ta}) \rangle.
\end{align*}
In the main text $\langle \varphi(x_\text{ta}) \rangle$ and $\langle \phi_d(x_\text{ta}) \rangle$ are displayed in Fig.~3, for $L=\SI{14}{\angstrom}$, which corresponds to the length of the thiophene backbone of the BTTT molecule.

An electronic state at energy $E_0$ (see Fig.~3a in the main text) appears, thus, in a tunneling spectrum at bias voltage:
\begin{equation}
 V^\ast_0 (x_\text{ta}) = \frac{E_0/e-\langle \phi_d (x_\text{ta})\rangle}{\langle \varphi(x_\text{ta})\rangle}
\end{equation}

\section{\label{App:CPD}Effect of CPD}
In the main text, we approximated the contact potential differece (CPD) with $V_\text{CPD}=0$.
In this section the effect of a non-vanishing CPD in the junction is discussed.
In good approximation, we can assume the total potential to depend linearly on the CPD \cite{NeffPRB2015,KocicJCP2017}:
\begin{equation}
  \phi_\text{tot} = \varphi \left ( \vec r \right ) \cdot \left ( V_\text{b} +V_\text{CPD} \right ).
\end{equation}

In case of an additional dipole at the tip, we have to expand the equation to
\begin{equation}
  \phi_\text{tot} \left ( \vec r \right ) = \varphi \left ( \vec r \right ) \cdot  V_\text{b} +\varphi \left ( \vec r \right ) \cdot  V_\text{CPD}  + \phi_\text{d} \left ( \vec r \right ).
\end{equation}

As can be seen, the bias-voltage-depended term is not affected by the CPD.
The bias-voltage-independent part $\phi_\text{static}$ on the other hand, now consists of both the dipole potential, as well as the CPD:
\begin{equation}
  \phi_\text{static} \left ( \vec r \right ) = \varphi \left ( \vec r \right ) \cdot  V_\text{CPD}  + \phi_\text{d} \left ( \vec r \right ).
\end{equation}
 
 Hence, the effect of a non-vanishing CPD can be viewed as a modification of the dipole term in our simple model.
The observation in Section~\ref{par:results} -- no effective dipole on the metal tip -- then implies that the CPD and the possible dipole field roughly cancel each other. 

\begin{acknowledgements}
We are grateful to the German research foundation for funding within the framework of the SFB 658 and TRR 227.
\end{acknowledgements}

\end{document}